\documentclass[12pt]{article}
\usepackage{amsmath, amssymb, amscd}

\begin{document}
\date{}

\title{Motion representation of one-dimensional cellular automaton rules}

\author{Nino Boccara\dag\ and Henryk Fuk\'s\ddag\\
\dag\ Department of Physics, University of Illinois, Chicago, USA\\
and\\
DRECAM/SPEC, CE Saclay, 91191 Gif-sur-Yvette Cedex, France\\
\texttt{boccara@uic.edu}\\
\ddag\  Department of Mathematics\\
Brock University\\
St. Catharines, Ontario L2S 3A1,\\
Canada\\
\texttt{hfuks@brocku.ca}}

\maketitle

\section*{Abstract}
Generalizing the motion representation we introduced for number-conserving rules, we give a systematic way to construct a generalized motion representation valid for non-conservative rules using the expression of the current, which appears in the discrete version of the continuity equation, completed by the discrete analogue of the source term. This new representation is general, but not unique, and can be used to represent, in a more visual way, any one-dimensional cellular automaton rule. A few illustrative examples are presented.

\section{Introduction} 
The cellular automaton (CA) model of highway car traffic proposed by Nagel and Schreckenberg~\cite{ns1992} uses a simple probabilistic cellular automaton rule to represent the motion of cars on a circular one-lane highway with neither entries or exits. Despite its simplicity, this model possesses quite a few of the essential characteristics of what is observed in a real highway traffic flow. This publication was rapidly followed by a huge number of papers on vehicular traffic (for a review see \cite{css2000}). In order to find all the possible highway CA traffic rules, we first published a paper~\cite{bf1998} in which we established a necessary and sufficient condition for a two-state CA rule to be number-conserving later followed by a second one~\cite{bf2002} in which was proved a more general theorem for $q$-state CA rules.\footnote{A possible generalization of this result would be to find a necessary and sufficient condition for CAs rule to be \emph{eventually number-conserving}. Although it is undecidable whether a one-dimensional cellular automaton obeys a given conservation law over its limit set, a variety of sufficient conditions to be satisfied by a one-dimensional CA to be eventually number-conserving have been recently established~\cite{b2006}.} Since the direction of motion described by number-conserving CA rules may vary in time, number-conserving cellular automata (CAs)  are more properly models of closed systems of interacting particles. When the number of states and number of inputs are not very small, the dynamics of the particles is, however, far from being clearly exhibited by the rule table. A simpler and more visual picture of the rule can be given by its \emph{motion representation}. The purpose of this paper is to define and give a systematic way to obtain such a representation we introduced for number-conserving CA rules~\cite{bf1998}. 

\section{Number-conserving CA rules}
A one-dimensional CA is a discrete dynamical system, which may be defined as follows. Let
$s:\mathbb{Z}\times\mathbb{N}\mapsto\mathcal{Q}$ be a function satisfying the equation
\begin{equation}
s(i,t+1)=f\big(s(i-r_\ell,t),s(i-r_\ell+1,t),\ldots,s(i+r_r,t)\big),
\label{evolution}
\end{equation}
for all $i\in\mathbb{Z}$ and all $t\in\mathbb{N}$, where $\mathbb{Z}$ denotes the set of all integers, $\mathbb{N}$ the set of nonnegative integers, and $\mathcal{Q}$ a finite set of states, usually
equal to $\{0,1,2,\ldots,q-1\}$. $s(i,t)$ represents the \emph{state of site $i$ at time $t$}, and the mapping
$f : {\mathcal{Q}}^{r_\ell+r_r+1}\to\mathcal{Q}$ is the \emph{CA evolution rule}. The positive integers $r_\ell$ and $r_r$ are, respectively, the \textit{left} and \textit{right radii} of the rule. In what follows, $f$ will be  referred to as either an \textit{$(r_\ell,r_r)$-radii CA rule} or an \textit{$n$-input CA rule},
where $n$ is the number $r_\ell+r_r+1$ of arguments of $f$. Following Wolfram~\cite{w1994}, to each rule $f$ we assign a \emph{code number} $N(f)$ such that
$$
N(f) = \sum_{(x_1,x_2,\ldots,x_n)\in{\mathcal{Q}}^n}
f(x_1,x_2,\ldots,x_n)q^{q^{n-1}x_1+q^{n-2}x_2+\cdots+q^0x_n}.
$$

When discussing family of specific CA rules it is usually convenient to divide the set of all one-dimensional $q$-state $n$-input  CA rules into \emph{equivalence classes} of rules having similar properties. These equivalence classes are defined with respect to the group generated by the operators of \emph{reflection} and \emph{conjugation} denoted, respectively by $R$ and $C$, and defined on the set of all one-dimensional $q$-state $n$-input CA rules $f$ by
\begin{align*}
R\,f(x_1,x_2,\ldots,x_n) & = f(x_n,x_{n-1},\ldots,x_1),\\
C\,f(x_1,x_2,\ldots,x_n) & = f(q-1-x_1, q-1-x_2,\ldots,q-1-x_n).
\end{align*}

\noindent\textbf{Definition} \textit{A one-dimensional $q$-state $n$-input CA rule $f$
is \textit{number-con\-serving} if, for all cyclic configurations of length
$L\ge n$, it satisfies
\begin{multline}
f(x_1,x_2,\ldots,x_{n-1},x_n)+f(x_2,x_3,\ldots,x_n,x_{n+1})+\cdots\\
+f(x_L,x_1\ldots,x_{n-2},x_{n-1})=x_1+x_2+\cdots+x_L.
\label{CRf}
\end{multline}}

In~\cite{bf2002} we proved the following theorem.

\vspace{0.2cm}

\noindent\textbf{Theorem} \textit{A one-dimensional $q$-state $n$-input CA rule $f$ is
number-con\-serving if, and only if, for all
$(x_1,x_2,\ldots,x_n)\in{\mathcal{Q}}^n$, it satisfies
\begin{align}
f(x_1,x_2,\ldots,x_n) = x_1 + \sum_{k=1}^{n-1}\big(
&f(\underbrace{0,0,\ldots,0}_k,x_2,x_3,\ldots,x_{n-k+1})\notag\\
-&f(\underbrace{0,0,\ldots,0}_k,x_1,x_2,\ldots,x_{n-k})\big),
\label{NScond}
\end{align}}

\vspace{0.2cm}

For number-conserving CA rules, the motion representation is defined as follows. List all the neighborhoods of an occupied site represented by its site value $s\in\mathcal{Q}$. Then, for each neighborhood, indicate the displacements of the $s$ particles by  arrow(s) joining the site where the particle(s) is (are) initially located to its (their) final position(s). A number above the arrow indicates how many particles are moving to the final position. To simplify the representation, we only list neighborhoods for which, at least one particle is moving. Here are two examples. The motion representation of the two-state three-input rule 184, which represents the simplest car moving rule on a one-lane highway, is
$$
\overset{\overset{1}{\curvearrowright}}{10}.
$$
Since a car located behind, that is, on the left of an occupied site cannot move, we do not mention it.  In a less compact notation we could mention this rule completing the representation above with
$$
\overset{\circlearrowleft}{1}1.
$$
The motion representation of the three-state three-input rule 6171534259461 is
$$
\overset{\overset{1}{\curvearrowright}}{10}\quad\overset{\overset{1}{\curvearrowright}}{11}\quad
\overset{\overset{2}{\curvearrowright}}{20}\quad\overset{\overset{1}{\curvearrowright}}{21}
$$
In both examples particles move only to the right.  It is, therefore, not necessary to indicate the state of the left neighboring site of the particle(s).

In \cite{ht1991} Hattori and Takesue  showed that a CA rule $f$ is
number-conserving if, and only if, for all 
$\{ x_1, x_2, \ldots, x_n \}\in \mathcal{Q}^n$ it satisfies
\begin{equation}
\label{curcons} 
f ( x_1, x_2, \ldots, x_n ) - x_{-r_\ell+1} = 
J ( x_1, x_2,\ldots, x_{n - 1} ) - J ( x_2, x_3, \ldots, x_n ),
\end{equation}
where
\begin{equation}
\label{curdef}
J(x_1,x_2,\ldots,x_{n-1}) = - \sum_{k=1}^{n-1}
f(\underbrace{0,0,\ldots,0}_k,x_1,x_2,\ldots,x_{n-k}) + \sum_{j=1}^{r_\ell} x_j.
\end{equation}
Equation~(\ref{curcons}) is actually a discrete version of  the continuity equation 
$\partial \rho / \partial t = - \partial J /\partial x$, where $\rho$ is the density of the conserved quantity and $J$ its current. The left-hand side of~(\ref{curcons}) is the discrete time derivative of the local particle density and its right-hand side represents minus the space derivative of the current $J$. 
More details regarding the current $J$ and its properties can be found in \cite{f2003,f2004}, and
for an extensive discussion of number-conserving rules refer to \cite{p2002, dfr2003,
fg2003}.

Here we shall just show how the current $J$ can be used to construct the motion representation of number-conserving rules.

For example, for $\mathcal{Q} = \{0,1\}$ and $r_\ell = r_r = 1$, 
apart the identity, there exist four number-conserving rules whose 
code numbers are 170 (left shift), 240 (right shift, reflected of 170), 
184 (mentioned above)  and 226 (reflected and conjugate of 184). 
For rule 184, the current $J_{184}$ is given by
$$
J_{184}(0,0)=0, \; J_{184}(0,1)=0,\; J_{184}(1,0)=1,\;J_{184}(1,1)=0,
$$
which can be simply written as
$$
J_{184}(x_1,x_2)=x_1 (1-x_2).
$$
The current is non-zero only for $(x_1,x_2)=(1,0)$, in agreement 
with the motion representation of rule 184 given above, namely,
$$
\overset{\curvearrowright}{10}.
$$  
The motion representation of rule 226, reflected of 184, is therefore
$$
\overset{\curvearrowleft}{01}.
$$
In the case of rule 170, we have
$$
J_{170}(0,0)=0,\; J_{170}(0,1)=-1,\;J_{170}(1,0)=0,\;J_{170}(1,1)=-1,
$$ 
or simply 
$$
J_{170}(x_1,x_2)=-x_2,
$$ 
and the corresponding motion representation is 
$$
\overset{\curvearrowleft}{\bullet1}.
$$
The motion representation of rule 240, reflected of 170, is therefore
$$
\overset{\curvearrowright}{1\bullet}.
$$

\section{Non-conservative CA rules}
For non-conservative rules, (\ref{curcons}) does not hold, 
but keeping the definition of  current $J$ given in (\ref{curdef}), we have to add 
a source term to the right-hand side of (\ref{curcons}) to take into account creations and 
annihilations of particles.  If we define this source term by
\begin{align*}
 \Sigma(x_1, x_2, \ldots, x_n ) =  & f ( x_1, x_2, \ldots, x_n ) - x_{-r_\ell+ 1} \\
 & - \big(J ( x_1, x_2, \ldots, x_{n - 1} ) - J ( x_2, x_3, \ldots, x_n )\big),
\end{align*}
the generalized continuity equation becomes
\begin{align}
 f ( x_1, x_2, \ldots, x_n ) - x_{-r_\ell + 1} = 
 J ( x_1, x_2, & \ldots, x_{n - 1} ) - J ( x_2, x_3, \ldots, x_n ) \notag\\
  &+ \Sigma(x_1, x_2, \ldots, x_n ).\label{curconsgen}
\end{align}
Depending upon its sign, positive or negative, the source term $\Sigma(x_1, x_2, \ldots, x_n )$ represents either creation or annihilation of particles. 

In order to illustrate how current and source terms can be used to construct
the motion representation, let us consider again the
case of elementary CA rules with  $r_\ell=r_r=1$. The
generalized continuity equation reads
\begin{equation}
   f ( x_1, x_2, x_3 ) - x_{2} = J ( x_1, x_2)
   - J ( x_2, x_3) + \Sigma(x_1, x_2, x_3),
\end{equation}
where
\begin{equation}
    J(x_1,x_2)= -f(0,x_1,x_2)- f(0,0,x_1) + x_1.
\end{equation}
For example, for monotone nonincreasing rule 56 we find
\begin{alignat*}{4}
J_{56}(0,0)&=0, &\quad J_{56}(0,1)&=0,&\quad
J_{56}(1,0)&=1,&\quad J_{56}(1,1)&=0,\\
\Sigma_{56}(0,0,0)&=0,&\quad \Sigma_{56}(0,0,1)&=0,&\quad
\Sigma_{56}(0,10)&=0,&\quad \Sigma_{56}(0,1,1)&=0,\\
\Sigma_{56}(1,0,0)&=0,&\quad \Sigma_{56}(1,0,1)&=0,&\quad
\Sigma_{56}(1,1,0)&=0,&\quad \Sigma_{56}(1,1,1)&=-1
\end{alignat*}
The current is the same as for rule 184, but, since $\Sigma_{56}(1,1,1)=-1$, the central 
particle is annihilated when $(x_1,x_2,x_3)=(1,1,1)$. We translate this fact 
in the motion representation by
$$
\overset{\curvearrowright}{10},\quad 1\overset{-1}{1}1,.
$$
In all other cases the particle remains at the same location. 
The full version of the motion representation of rule~56 is therefore
\begin{equation}
\bullet\overset{\curvearrowright}{10}, \quad1\overset{-1}{1}1,
 \quad 0\overset{\circlearrowright}{1}1.
\end{equation}

To describe the general procedure to construct the motion representation
from the expressions of $J$ and $\Sigma$ we consider the slightly more complicated 
case of rule 174. For this rule, the current is given by
\begin{equation*}
J_{174}(0,0)=0,\quad
J_{174}(0,1)=-1,\quad
J_{174}(1,0)=-1,\quad
J_{174}(1,1)=-1,
\end{equation*}
and the source term by
\begin{alignat*}{4}
\Sigma_{174}(0,0,0)&=0,&\quad
\Sigma_{174}(0,0,1)&=0,&\quad
\Sigma_{174}(0,1,0)&=0,&\quad
\Sigma_{174}(0,1,1)&=0,\\
\Sigma_{174}(1,0,0)&=1,&\quad
\Sigma_{174}(1,0,1)&=1,&\quad
\Sigma_{174}(1,1,0)&=-1,&\quad
\Sigma_{174}(1,1,1)&=0.
\end{alignat*}
Using the current definition, we create the motion representation by orienting the
arrow either to the right, when the current is equal to $1$, or to the left,when it is
$-1$. If the current is equal to $0$, we omit that entry. For rule 174, this yields
$$
\overset{\curvearrowleft}{01},\quad
\overset{\curvearrowleft}{10},\quad
\overset{\curvearrowleft}{11}.
$$
The middle term looks rather strange because 
a particle moves from an empty site to an occupied site,
but let us not worry about this now since the representation above is temporary. Taking
into account  the source term this anomaly will disappear. 
Note that the representation above can be written in a more compact form as
$$
\overset{\curvearrowleft}{\bullet1},\quad
\overset{\curvearrowleft}{10}.
$$
Now consider the creation-annihilation part. If $\Sigma_{174}=1$ (resp. $\Sigma_{174}=-1$), a particle 
is created (resp. annihilated) at the central site and this fact will be described writing $+1$ 
(resp. $-1$) above this site.  Entries for which $\Sigma_{174}=0$ are omitted.
Hence,
\begin{eqnarray*}
\Sigma_{174}(1,0,0)=1 \mbox{ yields } 1\overset{+1}{0}0,\\
\Sigma_{174}(1,0,1)=1 \mbox{ yields } 1\overset{+1}{0}1,\\
\Sigma_{174}(1,1,0)=-1 \mbox{ yields } 1\overset{-1}{1}0,
\end{eqnarray*}
Again, this can be written in a more compact form as
$$
1\overset{+1}{0}\bullet,\quad 1\overset{-1}{1}0.
$$
Combining entries obtained from the expressions of $J_{174}$ and $\Sigma_{174}$ 
we finally obtain
$$
\overset{\curvearrowleft}{\bullet1},\quad
\overset{\curvearrowleft}{10},\quad
1\overset{+1}{0}\bullet,\quad
1\overset{-1}{1}0.
$$
The third term implies the creation of a particle at any empty site located on the right of an occupied site, but, according to the second term, this particle will immediately move to its left neighboring site, the motion representation of rule 174 can be  written in a more compact way
$$
\overset{\curvearrowleft}{\bullet1},\quad
\overset{+1}{1}0,\quad
1\overset{-1}{1}0.
$$
This representation has to be interpreted carefully. The second term seems to imply that a particle is created at an already occupied site if it has an empty left neighboring site. This would result in a site occupied by two particles. It is, however, not the case since all particles move to the left, and the final result of the first and second terms is the creation of a particle at the first empty site located at the left end of a sequence of occupied sites. That is, 
$$
\overset{\curvearrowleft}{\bullet1}\;\;\text{and}\;\;\overset{+1}{1}0\;\;\text{is equivalent to}\;\;
{\overset{+1}0}1.
$$
We have thus obtained, for rule 174, the motion representation
$$
{\overset{+1}0}1, \quad 1\overset{-1}{1}0.
$$

\vspace{0.2cm}

\noindent\textbf{Remark} As illustrated by the considerations above, and mentioned in \cite{mgb2004}, for non-conservative rules, motion representations are not unique.

\section{Conclusion}
Using the notion of current, whose discrete divergence is related to the variation of the particle density in the discrete version of the continuity equation, completed by the definition of a source term, representing possible creations or annihilations of particles, we have shown how to derive from their expressions a generalized form of the motion representation, we introduced a few years ago to describe the temporal evolution of number-conserving cellular automaton rules. This new representation, which is not unique, shows in a more explicit way how, for each particular neighborhood, particles move or are either created or annihilated. A few illustrative examples have been given.

\end{document}